\documentstyle[epsfig]{aipproc}

\def\met{\mbox{${\hbox{$E$\kern-0.6em\lower-.1ex\hbox{/}}}_T$}} 
\def\gevcc{GeV/$c^2$}                   
\def\D0{D\O}                            

\def\Journal#1#2#3#4{{#1} {\bf #2}, #3 (#4)}

\def\NPB{{\em Nucl.~Phys.}~B}
\def\PLB{{\em Phys.~Lett.}~B}
\def\PRL{\em Phys.~Rev.~Lett.}
\def\PRD{{\em Phys.~Rev.}~D}

\def\be{\begin{equation}}
\def\ee{\end{equation}}
\def\bea{\begin{eqnarray}}
\def\eea{\end{eqnarray}}
\def\ba{\begin{array}}
\def\ea{\end{array}}

\def\simge{\mathrel{%
   \rlap{\raise 0.511ex \hbox{$>$}}{\lower 0.511ex \hbox{$\sim$}}}}
\def\simle{\mathrel{
   \rlap{\raise 0.511ex \hbox{$<$}}{\lower 0.511ex \hbox{$\sim$}}}}

\def\slashchar#1{\setbox0=\hbox{$#1$}           
   \dimen0=\wd0                                 
   \setbox1=\hbox{/} \dimen1=\wd1               
   \ifdim\dimen0>\dimen1                        
      \rlap{\hbox to \dimen0{\hfil/\hfil}}      
      #1                                        
   \else                                        
      \rlap{\hbox to \dimen1{\hfil$#1$\hfil}}   
      /                                         
   \fi}                                         %
\def\ts{\thinspace}

\def\ra{\rightarrow}

\def\CO{{\cal O}}

\def\atro{\alpha_{\rho_T}}

\def\kslash{\raise.15ex\hbox{/}\kern-.57em k}
\def\tro{\rho_{T}}

\def\troz{\rho_{T}^0}

\def\tom{\omega_T}
\def\tpi{\pi_T}

\def\tpimp{\pi_T^\mp}

\def\tpiz{\pi_T^0}
\def\tpipr{\pi_T^{0 \ts\prime}}

\def\mm{\mu^+\mu^-}

\def\gev{{\rm GeV}}

\def\nb{{\rm nb}}
\def\pb{{\rm pb}}

\begin{document}

\title{
\begin{flushright}
{\small\textrm FERMILAB-conf-98/072}
\end{flushright}
Strong Dynamics at the Muon Collider:\\
Working Group Report\footnote{Summary talk presented by P.C. Bhat at
``Workshop on Physics
at the First Muon Collider and at the Front End of a Muon Collider'',
November 6-9, 1997, Fermilab, Batavia, IL 60510}}

\author{
Pushpalatha C. Bhat\footnote{Co-convenors} and
Estia Eichten$^{\dagger}$
}
\vspace{-3.0cm}
\address{
Fermi National Accelerator Laboratory,
Batavia, IL 60510 \\
\vspace{1.0cm}
}

\maketitle

\begin{center}
{\small Working group members:
Gustavo Burdman,
Bogdan Dobrescu,\\
Daniele Dominici,
Takanobu Handa,
Christopher Hill,
Stephane Keller,\\
Kenneth Lane,
Paul Mackenzie,
Kaori Maeshima,
Stephen Parke,\\
Juan Valls,
Rocio Vilar and
John Womersley}
\end{center}

\begin{abstract}
New strong dynamics   at  the energy  scale  $\approx$  1  TeV  is an
attractive   and   elegant theoretical   {\it ansatz}   for  the origin
of electroweak symmetry breaking.  We review here, the theoretical models
for strong dynamics,  particularly, technicolor theories and their low
energy signatures.   We  emphasize  that  the  fantastic   beam energy
resolution ($\sigma_E  /E \sim ~10^{-4}$) expected   at the first muon
collider   ($\sqrt{s}$  =  100--500~GeV)   allows the   possibility of
resolving  some  extraordinarily narrow  technihadron  resonances and,
Higgs-like techniscalars produced   in  the s-channel.   Investigating
indirect   probes   for strong  dynamics   such    as search for  muon
compositeness, we find that  the muon colliders provide unparallel
reaches. A big muon collider (${\sqrt s}$ =3--4~TeV) would be a remarkable
facility  to study heavy  technicolor  particles such as the  topcolor
Z$^\prime$, to   probe the dynamics     underlying fermion masses  and
mixings and  to  fully  explore the strongly   interacting electroweak
sector.
\end{abstract}

\section*{Introduction}

The   success of  the Standard  Model   of Particle  Physics  has been
spectacular, thus far! But, new physics beyond the Standard Model (SM)
seems  inevitable since some critical  issues  remain unresolved.  The
cause  of electroweak symmetry breaking  (EWSB)  is not experimentally
established,  nor is the origin  of fermion  masses and mixings known.
There  are two enticing theoretical approaches
to understand EWSB -- introducing  supersymmetry or invoking new strong
interactions.  Our working group explored the latter scenario, that of
a new strong  dynamics, and, its  search and study  at the First  Muon
Collider.

In this report, we review and summarize  the activities of the working
group.  We first explore the existing
theories for new strong  dynamics.  These models 
provide an intuitively attractive (though presently disfavored)    
approach to origin of EW symmetry breaking.
A large fraction  of this report is devoted to
topics related to technicolor, topcolor and their variants.
We
give an  overview  of  the   technicolor  models, their   low   energy
signatures, the potential of the  First Muon Collider (FMC with $\sqrt
s$=  100-500  GeV) for direct  searches  and detailed  measurements of
these signatures.  Some comments on  how these compare with what would
be attainable at  the Tevatron, LHC  and the possible NLC machines are
included.  We  also explore the  indirect  probes for strong  dynamics
such as tests for compositeness
and comment on constraints from rare B and K decays. We briefly state the
long-range opportunities to  discover  and study new  strong  dynamics
with a big muon collider (BMC with $\sqrt s$ = 3--4~TeV).

At  the working  group  meetings,  technicolor theories~\cite{tc}  and
relevant issues  for the FMC  were described  in  detail by  Ken  Lane
~\cite{lane}.         The    specific   details  for topcolor
theories~\cite{topc} and other   recent new ideas  in strong  dynamics
were expounded  by   Chris Hill  ~\cite{Hill}.   Cross  sections   for
production of some low mass technihadrons at a $\mu ^+\mu ^-$ collider
and  their signatures were discussed  by  John Womersley ~\cite{JW}. A
model  of technicolor with scalars   and the prospects for discovering
non-standard  Higgs-like scalars  in   the s-channel at  the  FMC were
presented by Bogdan Dobrescu~\cite{bdob}. A study of vector resonances
in  the  framework of  BESS   (breaking electroweak symmetry strongly)
model~\cite{bess}  was  presented by Daniele Dominici~\cite{dominici}.
A search  for technicolor particles using  the Tevatron Run I data and
the  resulting   95\%  confidence level (C.L.)    upper  limits on the
production cross-section and exclusion of certain mass regions for
these  particles were reported by the  CDF
collaboration  ~\cite {cdftc}.  Studies   of various indirect tests of
strong dynamics   such   as compositeness  tests   (Eichten and Keller
~\cite{keller}),  strong WW   scattering (Gunion ~\cite{gunion})   and
constraints on strong  dynamics  from rare   B and K   decays (Burdman
~\cite{burdman}), were also presented.   During the workshop, prompted
by comments from Hill  and   Mackenzie, stressing the  importance   of
narrow neutral technipion production at  a muon collider, Eichten  and
Lane calculated  the cross sections  for resonance production of these
particles  at  the  FMC.  Subsequently,  Dominici   et al., have  also
studied   the     production    of     such       particles    (called
pseudo-Nambu-Goldstone  bosons -- PNGBs) in  the framework of the BESS
model~\cite{dominici}.

\section*{Technicolor and Variants}
{\bf Technicolor (TC)} is  a new strong interaction of
fermions and gauge bosons
at the scale $\Lambda_{TC}\sim$ 1 TeV, which causes dynamical breaking
of   electroweak symmetry~\cite{tc}. No elementary   scalar  bosons (such as the
Higgs)  are required.  Technicolor  model, in its  simplest form, is a
scaled-up version  of QCD with  massless  technifermions that strongly
interact at a scale $\Lambda_{TC} \sim$  1 TeV and acquire a dynamical
mass  ${\cal O}(\Lambda_{TC})$.  The  chiral symmetry is spontaneously
broken through technifermion  condensation,  producing three  massless
Goldstone bosons. These Goldstone bosons (technipions) have Higgs-like
coupling to fermions  and  correspond to the  longitudinal  components
$W_L^\pm$ and   $Z_L^0$   of the  weak  gauge  bosons.  If  left-  and
right-handed technifermions  are assigned to   weak SU(2) doublets and
singlets,   respectively,  then  $M_W=M_Z  \cos\theta_W={\frac{1}{2}}g
F_\pi$   where  $F_ \pi$=246 GeV is    the  technipion decay constant,
analogous  to   $f_\pi$  =  93 MeV   for   the  pion.   In non-minimal
technicolor  model, with a   large  number of technifermion  doublets,
additional Goldstone bosons arise  from technifermion  chiral symmetry
breaking.  The  technicolored  and the  SM  fermions  however remain
massless. They can acquire masses  if they couple to technifermions via
additional gauge interactions as shown  in Fig~\ref{fig:etc}. In  this
{\bf Extended Technicolor (ETC)}  model~\cite{etc},  
the quark and lepton  masses
($m_f$)  proportional to  the   dynamical mass of  the  technifermions
(condensate $<T\bar{T}>$) are generated:

\begin{figure}
\vspace*{6cm}
\includegraphics{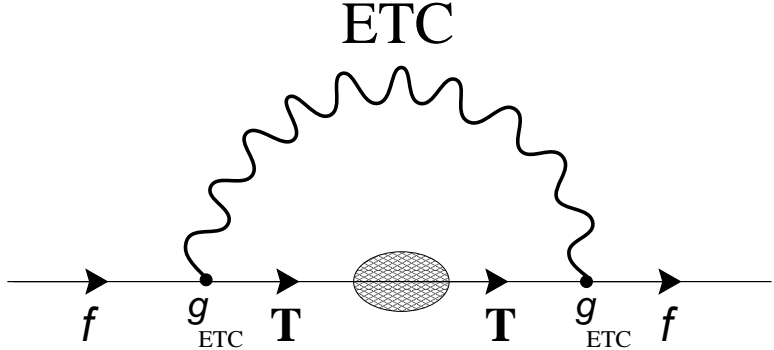}
\caption[etc]{Generation of quark and lepton masses 
via ETC interactions.}
\label{fig:etc}
\end{figure}

\begin{equation}
$$m_f(M_{ETC}) \approx {\frac
{g^2_{ETC}}{M^2_{ETC}}}<T\bar{T}>_{ETC}$$
\end{equation}
where $g_{ETC}$ is  the coupling strength of the  fermions to  the ETC
boson  and $M_{ETC}$ is the  mass of the ETC  boson.  The ETC symmetry
which  pertains to a larger gauge  group into which technicolor, color
and     flavor symmetries  are  embedded,     is   broken at  a  scale
$\Lambda_{ETC}={\cal O}$(100 TeV).

To avoid large flavor-changing neutral   currents and to obtain  quark
masses of   a few GeV,  the  strong technicolor coupling $\alpha_{TC}$
must run very slowly or  ``walk'', all the way  up to the ETC scale  of
several hundred TeV~\cite{wtc}.   
{\bf Walking Technicolor}  needs a large number
of technifermions for $\alpha_{TC}$ to ``walk''.

Another major turning point in the development of technicolor theories
came with the discovery  of the top quark and  the measurement  of its
large mass ($m_t$) ~\cite{cdfobs,d0obs}.  The direct measurements from
the CDF and D\O\ collaborations yield a value of $m_t = 175.6 \pm$ 5.5
\gevcc\  (current  world average) ~\cite{cdfmass,d0mass}.  To generate
this large  $m_t$, the ETC models  would have  to violate experimental
constraints   on  the     $\rho$  parameter  ($\rho=\frac    {M_W}{M_Z
cos\theta_W}$) or the Z  $\rightarrow b\bar{b}$ decay rate. To resolve
this problem, a new strong  {\bf Topcolor} interaction was introduced by
Hill~\cite{Hill} and {\bf  Topcolor-assisted Technicolor}  ($TC2$) was
born~\cite{tctwo}.  
The top quark is very heavy compared to all the other quarks and leptons.
This fact suggests that it might be strongly coupled  to the mechanism of
mass generation and to the dynamics of  EWSB itself. 
It is  conceivable that the top quark  has unique dynamics. 
The simplest $TC2$
model has the following group structure:

\be
\ba{c}
G_{TC}
\times SU(2)_{EW} \times  SU(3)_3 \times SU(3)_{1,2}
 \times U(1)_3 \times U(1)_{1,2} \\
\longrightarrow G_{TC} \times  SU(2)_{EW} \times SU(3)_C   \times
U(1)_Y \\
\longrightarrow SU(3)_C \times  U(1)_{EM}
\ea
\ee
where  $G_{TC}$ and $SU(2)_{EW}$ are   the technicolor and electroweak
gauge groups; $SU(3)_3$ and $U(1)_3$ are topcolor gauge groups coupled
to the third generation  fermions  (with stronger couplings)  while  $
SU(3)_{1,2}$ and $U(1)_{1,2}$  couple to first  and second generations
only.  Technicolor causes most of the EWSB, while topcolor contributes
only  feebly with  $f_t \approx$ 60  GeV. The  light  quark and lepton
masses are generated via ETC dynamics  which contributes only a GeV to
the third generation masses.  The  strong topcolor dynamics (top quark
pair  condensate) generates   $m_t   \approx$ 175  GeV. The   $U(1)_3$
provides the difference that causes only top quarks to condense. Thus,
the  top   quark  mass may  be   perceived as   being generated  by  a
combination   of a  dynamical condensate component,  $(1-\epsilon)m_t$
(from topcolor dynamics) and  a small fundamental component, $\epsilon
m_t$ (from {\it e.g.,} technicolor) with $\epsilon  <<1$.  A number of
additional  particles  called  ``top-pions''  $\pi_t$ and  ``top-rho''
$\rho_t$,  are expected.  The  small ETC  component  of the top  quark
implies that the masses   of the top-pions  depend on   $\epsilon$ and
$\Lambda$.  The  top-pion  mass induced from  the  fermion loop can be
estimated as,
\be
M_{\pi_t}^2 = \frac {N\epsilon  m_t^2 M_B^2}{8\pi^2 f_{\pi}^2} = \frac
{\epsilon M_B^2}{log (M_B /m_t)}
\ee
where the Pagels-Stoker   formula is   used  for $f_{\pi}^2   $.   For
$\epsilon$  = (0.03, 0.1), $M_B \approx   $ (1.5, 1.0)~TeV, and $m_t$=
180 GeV, this predicts $M_{\pi_t}$ = (180,  240) GeV.  The bare values
of $\epsilon$  generated    at $\Lambda_{ETC}$ is    subject  to large
radiative enhancements ($\sim$10) by topcolor and $U(1)_{3}$.  Thus,
we expect that even a bare value  of $\epsilon \sim $0.005 can produce
sizeable   $M_{\pi_t}$ ($>m_t$).   The  breaking   of  $U(1)_3  \times
U(1)_{1,2} \rightarrow U(1)_Y$ in the vicinity of 2 TeV leads to eight
color-octet vector bosons $V_8$ or $B$ (colorons or top-gluons) and an
additional  Z boson, Z$'$. The  mass of the Z$'$  is expected to be in
the range of 1-3 TeV.

\subsection*{Top See-saw Model}
The topcolor models have met with  some problems in their implications
for limits on custodial symmetry  violation and other phenomenological
constraints.  The proximity of the measured top quark mass, $m_t $, to
the electroweak scale, however, suggests that EWSB may have its origin
in dynamics associated with the top quark.  An explicit realization of
this  idea is the top  condensation mechanism \cite{topcond}, in which
the  top-antitop quark pair acquires  a vacuum expectation value, much
like the  chiral  condensate  of QCD  or   the electron condensate  of
superconductivity (BCS theory).  The EWSB  occurs via the condensation
of   the   top  quark  in   the presence   of   an extra  vector-like,
weak-isosinglet   quark.  The   mass    scale of  the    condensate is
$\sim$0.6~TeV  corresponding  to  the  electroweak  scale   $v\approx$
246~GeV.  The vector-like isosinglet then naturally exhibits a see-saw
mechanism,  yielding the   physical  top quark   mass,   which is then
adjusted to the experimental value.  The choice of $\sim$TeV scale for
the  topcolor   dynamics determines the  mass   of  the weak-isoscalar
see-saw  partner.  The model also implies  the existence of PNGBs. The
lower bound on  the mass of a  PNGB that couples  to the top  quark is
less than $m_t$.

More work is needed to extend the scheme to generate masses and mixing
for  all quarks and leptons,  and  to construct attractive schemes for
topcolor breaking.

\subsection*{Technicolor Production and Signatures at the FMC}

In the minimal technicolor model, with just one technifermion doublet,
the only prominent signals at the hadron and lepton colliders would be
the   enhancements  in    longitudinally-polarized weak  vector  boson
production.  These   are    due  to  the   s-channel   production   of
color-singlet technirho resonances near 1.5-2.0 TeV and the subsequent
decay into vector  boson pairs ($\rho_T^0 \rightarrow W_L^+W_L^-$  and
$\rho_T^{\pm}       \rightarrow   W_L^{\pm} Z^0$).   Observing   these
enhancements   would  be extremely difficult,    since  the ${\cal O}
(\alpha^2)$ production cross sections are small at such high technirho
masses and efficiency for reconstructing vector boson pairs low.

The non-minimal technicolor  models, however, predict  a rich spectrum
of   light,    color-singlet  technihadrons---the  isotriplet  vectors
$\rho_T^0,  \rho_T^{\pm}$ and their  isoscalar partner $\omega_T$, and
pseudoscalars  $\pi_T^0,  \pi_T^{\pm}$ and $\pi_T^{0}{'}$---accessible
at the Tevatron, LHC and the FMC. (A search at the Tevatron by CDF has
been discussed later).  Since techni-isospin is likely   to be a  good
approximate   symmetry,  $\rho_T$   and  $\omega_T$ are  approximately
degenerate and so are the technipions. The  masses are expected to be:
$m_{\pi_T}     \approx    100~GeV$       and   $m_{\rho_T}     \approx
m_{\omega_T}\approx  200$~GeV.   The technipions with  Higgs-like  ETC
couplings  to quarks and leptons decay   to the heaviest fermion pairs
allowed.   The   isosinglet   component     of   neutral  technipions,
$\pi_T^{0}{'}$, may decay into  a  pair of  gluons if the  constituent
technifermions  are colored. Thus the  predominant decay signatures of
the light technipions would be:

\begin{equation}
\begin{array}{c}
$$\pi_T^0 \rightarrow b\bar{b},~c\bar{c}, ~\tau^+\tau^- \\
\pi_T^{0}{'} \rightarrow gg,~b\bar{b},~c\bar{c}, ~\tau^+\tau^- \\
~\pi_T^+
\rightarrow c\bar{b}, ~c\bar{s}, ~\tau^+\nu_{\tau}.$$
\end{array}
\end{equation}

The signatures for technirhos and techniomegas are as follows:

\begin{equation}
\begin{array}{c}
$$\rho_T^{\pm} \rightarrow W^{\pm} Z,W^{\pm}\pi_T^0,
Z \pi_T^{\pm}, \pi_T^{\pm} \pi_T^0  \\
\rho_T^0 \rightarrow
W^{+} W^{-},~W^{\pm}\pi_T^{\mp},~ \pi_T^{\pm} \pi_T^{\mp},~q\bar{q},
~ \ell^+\ell^-,~ \nu \bar{\nu} \\
\omega_T \rightarrow \gamma \pi_T^0, ~Z \pi_T^0, ~q\bar{q},
~\ell^+\ell^-,~\nu \bar{\nu}.$$
\end{array}
\end{equation}
If the  large  ratio of $\frac   {<\bar{T}T>_{ETC}} {<\bar{T}T>_{TC}}$
significantly enhances technipion   masses  relative  to  technivector
masses, then $\rho_T  \rightarrow \pi_T \pi_T$  decay channels  may be
closed.

If technicolor exists  and technihadrons have masses  low enough to be
produced at  the FMC, they  will most probably  be first discovered at
the Tevatron or at the LHC. An interesting aspect of the technihadrons
is that  several  of  them, particularly the   neutral ones  are  very
narrow. Therefore, a $\mu^+\mu^-$  collider which is expected  to have
very fine energy resolution is ideally suited for their studies.

\begin{figure}[t] 
\begin{minipage}[t]{2.5in}
\vspace*{6.5cm}
\includegraphics{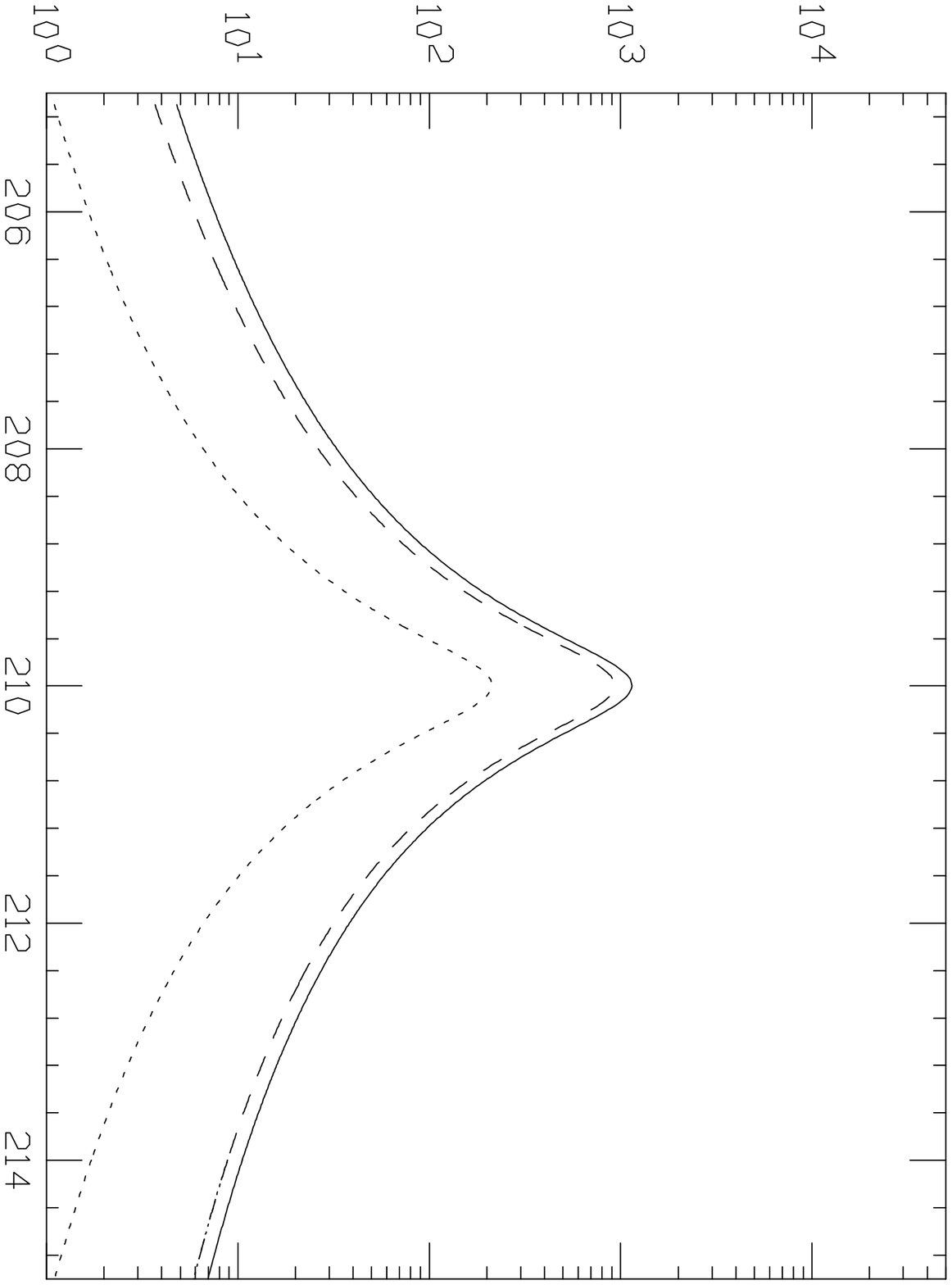}
\caption[womer1]{Cross   section  (pb) for   technirho  production  at  a muon
collider as a function of  ${\sqrt s}$(GeV), for  $M_{\rho_T}$=210~GeV
and  $M_{\pi_T}$=100~GeV. The solid curve is  the total cross section,
the dashed curve  is for $\rho_T \ra  W\pi_T$ and the dotted curve  is
for $\rho_T \ra W^+W^-$.}
\label{fig:womer1}
\end{minipage}\hfill
\begin{minipage}[t]{2.5in}
\vspace*{6.5cm}
\includegraphics{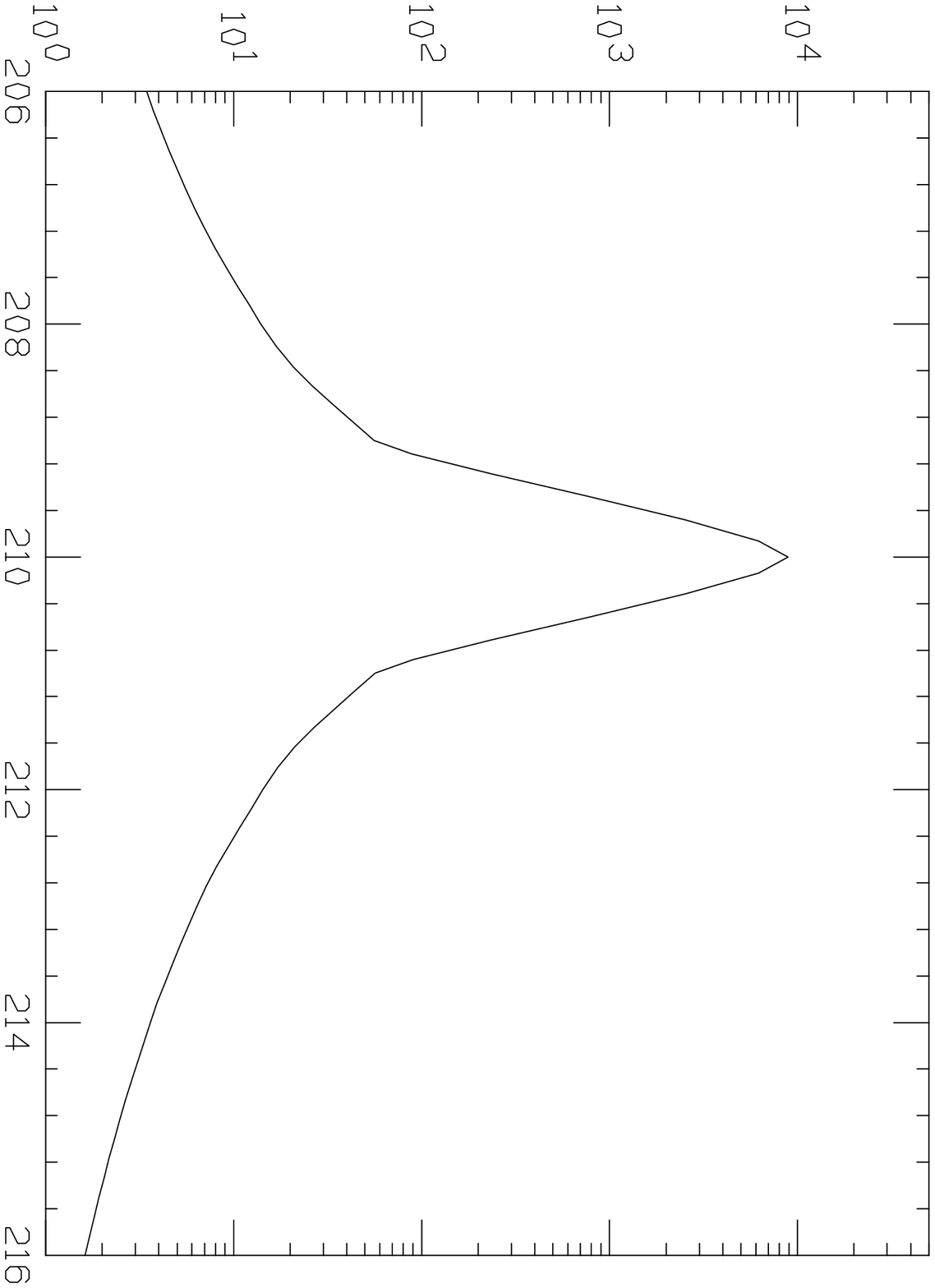}
\caption[womer2]{Total cross  section  (pb) for   techniomega production  
at  a muon
collider as a function of ${\sqrt s}$(GeV), for $M_{\omega_T}$=210~GeV
and $M_{\pi_T}$=110~GeV. The dominant decay mode is $\gamma\pi_T^0$.}
\label{fig:womer2}
\end{minipage}\hfill
\end{figure}

Figure  ~\ref{fig:womer1}  shows  the  production  cross  section  for
$\rho_T$   at  a  muon collider  as   a function  of   $\sqrt  s$, for
$M_{\rho_T}$=210 GeV and $M_{\pi_T}$=110  GeV. The peak cross  section
is $\sim$ 1 nb which translates to $10^6$ events/year with $\int {\cal
L}dt$=$10^{32}~cm^{-2}~s^{-1}$.  Figure   ~\ref{fig:womer2}  shows the
cross section for $\omega_T$  production ($M_{\omega_T}$=210 GeV). The
peak  cross section here  is even  larger,  $\sim$ 10  nb,  that would
provide an yield of $10^7$  events/year for the same luminosity. Also,
note that the peak is  extremely narrow, with a  width $<$ 1 GeV.  The
production  cross   sections   decrease,  if $\rho_T,~~\omega_T$   are
heavier. For $M_{\rho_T}$=400  GeV and $M_{\pi_T}$=150 GeV, the  event
rate is still $10^4$ events/year.

 The neutral technipions,  like the  SM Higgs  boson, are  expected to
couple to  $\mu^+\mu^-$ with a  strength proportional to $m_{\mu}$. In
the  {\it ansatz} of   the  non-minimal technicolor model with   $N_D$
technifermion   doublets,  the coupling is  enhanced   by a  factor of
$\sqrt{N_D}$.  Therefore, the  FMC can serve   as a neutral technipion
factory with phenomenal rates for   production in the s-channel,   far
exceeding those at any other collider.

\begin{figure}[t]  
\vspace*{8cm}
\includegraphics{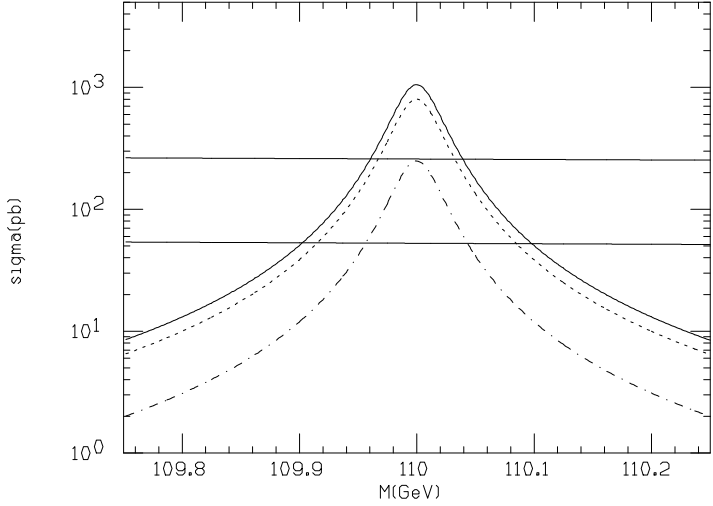}
\caption[ken1]{Theoretical   (unsmeared)  cross sections  for  $\mu^+\mu^-
\rightarrow     \pi_T^0{'} \rightarrow    b\bar{b}$   (dashed),   $gg$
(dot-dashed)  and total (solid) for  $M_{\pi_T}=  110\,\gev$ and other
parameters defined  in the text.   The solid horizontal  lines are the
backgrounds  from $\gamma, Z^0 \rightarrow  b\bar{b}$ (lower) and $Z^0
\rightarrow q\bar{q}$ (upper). Note the energy scale.  }
\label{fig:ken1}
\end{figure}

 Once a  neutral technipion has been found   in $\rho_T$ or $\omega_T$
decays at a  hadron collider, it should be  relatively easy  to locate
the precise position of the resonance at the FMC and, take data at the
resonance. The cross sections for $f\bar{f}$ and $gg$ final states are
isotropic.  The   $\tpipr$  production cross  sections and   the $Z^0$
backgrounds  are shown  in Fig.~\ref{fig:ken1} for  $M_{\tpi} = 110\,\gev$
(for description of other  parameters see ref.~\cite{lane}).  The peak
signal rates approach $1\,\nb$.   The $b\bar{b}$ dijet rates are  much
larger than the $Z^0 \ra b\bar{b}$ backgrounds, while the $gg$ rate is
comparable   to $Z^0  \ra   q\bar{q}$.   Details  of these   and other
calculations in this section, including the effects of the finite beam
energy resolution, will appear in Ref.~\cite{elwfmc}.

The  cross  sections for   technipion  production   via the decay   of
technirho and techniomega  $s$-channel resonances are calculated using
vector                  meson           ($\gamma$,              $Z^0$)
dominance~\cite{ehlq,multi,tpitev,elwtev}.  For $M_{\tro} = M_{\tom} =
210\,\gev$, $M_{\tpi} = 110\,\gev$, and other parameters as above, the
total peak cross sections are~\cite{elwfmc}:

\be
\ba{ll}
&\sum_{AB} \ts \sigma(\mm \ra \troz \ra \pi_A \pi_B)  = 1.1\,\nb \\ \\
&\sigma(\mm \ra \tom \ra \gamma \tpiz)  = 8.9\,\nb \ts.
\ea
\ee
%


The technirho decay rate is 20\% $W^+W^-$ and 80\% $W^\pm \tpimp$.

\begin{figure}
\vspace*{8cm}
\includegraphics{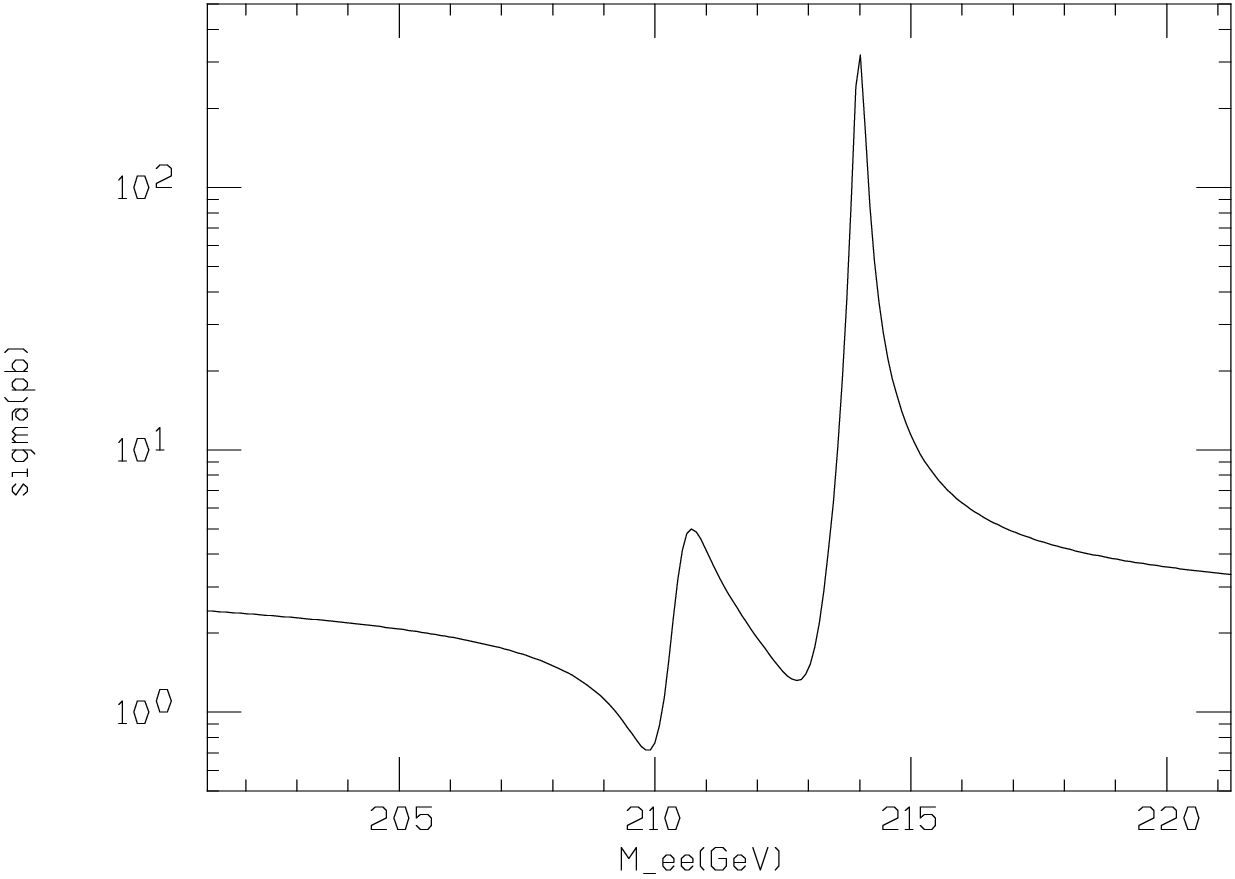}
\caption[ken2]{  Theoretical (unsmeared) cross  sections for $\mm \ra \troz,
\tom \ra e^+e^-$ for input masses $M_{\tro} = 210\,\gev$ and $M_{\tom}
= 212.5\,\gev$ and other parameters as defined in the text.  }
\label{fig:ken2}
\end{figure}

Further,  there  might be a  small  nonzero  isospin splitting between
$\troz$ and $\tom$.  This would appear as  a dramatic interference  in
the  $\mm  \ra  f\bar{f}$   cross  section, provided the   FMC  energy
resolution is good  enough  in the  $\tro$--$\tom$  region. The  cross
section is most  accurately calculated~\cite{lane}  by using the  full
$\gamma$--$Z^0$--$\tro$--$\tom$ propagator matrix ($\Delta$).

Figure~\ref{fig:ken2} shows  the theoretical $\troz$--$\tom$  interference
effect in $\mm \ra e^+e^-$ for input masses $M_{\tro} = 210\,\gev$ and
$M_{\tom} = 212.5\,\gev$. The propagator  shifts the nominal positions
of the resonance peaks  by  $\CO(\alpha/\atro)$. The theoretical  peak
cross sections  are $5.0\,\pb$   at  $210.7\,\gev$ and  $320\,\pb$  at
$214.0\,\gev$.  This  demonstrates   the  importance    of   precision
resolution in the $200\,\gev$ FMC.

The detectors~\cite{lebrun} at the muon  collider should be capable of
identifying and measuring electrons, muons, taus, jets and, of tagging
$b$-jets with high efficiency. It would be useful if $c$-jets could be
distinguished from $b$-jets.

\subsection*{Topcolor Signatures}
Topcolor-assisted  technicolor  introduces additional particles called
top-pions ($\pi_{t})$, top-gluons (B  or $V_8$) and topcolor  Z$'$, as
discussed in the   previous section.  Top-pions  can  be  as light  as
$\sim$150~GeV,  in which   case  they would  emerge as   a  detectable
branching  fraction of top   quark  decay.  However,  not to   violate
constraints on  $Z\rightarrow b\bar{b}$ rate, $M_{\pi_t}  \ge $300~GeV
may be required. Top-gluons are expected to have mass  in the range of
0.5-2~TeV and topcolor  $Z'$ in the range of  1-3~TeV.  The decays are
expected to be:

\begin{equation}
\begin{array}{c}
$$\pi_t \rightarrow t\bar{b},\quad \rm{or} \quad t \rightarrow \pi_{t}b \\
B \rightarrow b\bar{b}, t\bar{t}  \\
Z' \rightarrow t\bar{t}.$$
\end{array}
\end{equation}
Top-pions may  be produced copiously at  the FMC in the $s$-channel as
previously  discussed in the case  of technipions. The LHC experiments
should  be sensitive over the entire  range of the expected masses for
both top-gluons and topcolor $Z'$.   If topcolor $Z'$  is not found at
LHC,  it can be  discovered  at the big    muon collider ($\sqrt  s$ =
3--4~TeV).   There are a number of  other effects of topcolor that can
be observed at  the FMC~\cite{Hill}.  For example,  new effects in $Z$
physics involving the third generation such  as $Z\rightarrow b\bar b$
, might be   observed.  The  generational  structure of  topcolor  may
induce GIM violation in low  energy processes such as $K^+ \rightarrow
\pi^+ \nu \bar {\nu}$ and lepton  family number violation such as $\mu
\bar{\mu}\rightarrow \tau \bar {\mu}   $.  There may be  induced  FCNC
interactions giving rise to anomalous $\mu \bar{\mu}\rightarrow b \bar
s $.  The FMC and the front-end of the FMC provide great opportunities
to study such effects that  are enhanced due  to topcolor {\it w.r.t.}
SM.

\subsection*{Technicolor with Scalars}

Technicolor models that include scalars   are an interesting class  of
models for dynamical     EWSB. In the  current  model~\cite{bdob},  in
addition  to SM fermions, one doublet  of technifermions, $P$ and $N$,
and   three scalars, $\phi$, $\chi$   and  $\Phi$ are considered.  The
gauge  group      considered   is   $SU(4)_{TC}\times    SU(3)_C\times
SU(2)_{EW}\times U(1)_Y$.  Only the third   generation couples to  the
technicolor fields,  and,   as in QCD, the    $SU(4)_{TC}$ techincolor
interactions  trigger  the  formation  of   technifermion  condensates
$<P\bar{P}>\simeq<N\bar{N}>\simeq 2  {\sqrt  3}\pi f^3$, which  breaks
the electroweak symmetry  at a scale  $f$.   This also  results in the
generation of masses for $t,b$ and $\tau$. The masses of the first and
second generations  are generated by coupling  to a scalar $\Phi$ that
behaves like a Higgs  doublet under gauge transformations.  The scalar
acquires a small vacuum expectation value (VEV) by coupling to the new
strong   interactions sector and would  have  Yukawa  couplings to the
first and second generations which are larger than in the SM.

If this model is the correct description of physics up to a TeV scale,
then  the components of  the $\Phi$ scalar  should  be accessible at a
$\mu^+\mu^-$ collider  with  $\sqrt s$  below  the first  technihadron
resonance. Since the Yukawa  couplings are proportional to the fermion
mass, the s-channel production is very large  at a muon collider.  The
scalar $\Phi$   decomposes   into  an   isosinglet  $\sigma$  and  an
isotriplet $\pi'^3$, $a$=1,2,3. The neutral  real scalar $\sigma$ and
the charged   scalars  $\pi'^{\pm}={\frac {(\pi'^1\mp  i\pi'^2)}{\sqrt
2}}$ are almost degenerate, with a mass $M_\Phi$.  For $M_\Phi < \sqrt
s < 2M_\Phi$, only the $\sigma$ and $\pi{'}^3$, can be produced.

The total decay widths of the $\sigma$ and $\pi{'}^3$ scalars are equal.
The VEV of $\Phi$ is taken to be in the range,

\begin{equation}
$$1GeV \leq f{'} \leq 10GeV$$,
\end{equation}
where the  lower  bound is chosen to   avoid Yukawa coupling constants
larger than  1.0,   and upper  bound  is chosen  to satisfy  condition
$f'<<f$.  The  width     for decay  into   pairs   of  gauge   bosons,
$\Gamma(W^+W^-+ZZ)$, is    at most a  few  percent   of the width  for
$\sigma,\pi{'}^3 \rightarrow  c\bar{c}$,  and  is neglected here.  The
widths  of the $\sigma$ and  $\pi{'}^3$ scalars are  dominated only by
the $c\bar{c}$ final state:

\begin{equation}
\Gamma \approx {\frac{3m_c^2M_\Phi}{8\pi f{'}^2}} \approx 13.2 GeV \biggl(\frac
{3GeV}{f{'}}\biggr)^2 \biggl(\frac{M_\Phi}{500 GeV}\biggr)^2.
\end{equation}

Given the  enhanced couplings to the  second generation, the s-channel
production  of the   neutral scalars  at  a  $\mu^+\mu^-$  collider is
large.   The natural   spread   in  the  muon  collider   beam energy,
$\sigma_{\sqrt{s}}$, is rather small, and  can be ignored in computing
the effective s-channel resonance cross section:

\begin{equation}
\bar{\sigma}(\mu^+ \mu^- \ra \sigma,\pi{'}^3 \rightarrow X)\approx
{\frac{4\pi\Gamma ^2} {(s - M_{\Phi}^2)^2 +  
M_{\Phi}^2\Gamma^2}}B(\sigma,\pi{'}^3 \rightarrow
\mu^+\mu^-)B(\sigma,\pi{'}^3 \rightarrow X).
\end{equation}

 For the final state is $X \equiv c\bar{c}$, this cross section becomes,

\begin{equation}
\bar{\sigma}(\mu^+ \mu^- \ra \sigma,\pi{'}^3 \rightarrow c\bar{c})\approx  
{\frac{4\pi\Gamma ^2}
{(s - M_{\Phi}^2)^2 + M_{\Phi}^2\Gamma^2}}\biggl(\frac{m_{\mu}^2}{3m_c^2}\biggr).
\end{equation}

The main background comes from $\mu^+\mu^- \rightarrow \gamma^*,Z^*
\rightarrow c\bar{c}$, and amounts to

\begin{equation}
\sigma(\mu^+\mu^- \rightarrow c\bar{c})\approx 0.7pb{\frac{(500GeV)^2}{s}}.
\end{equation}

The  discovery potential of  a  $\mu^+\mu^-$ collider  operating at  a
maximum center of mass  energy of 500  GeV has been studied. Two  scan
points, at 300 and 500 GeV, are sufficient to find the neutral scalars
with masses roughly between 200 and 600 GeV.

Once the  resonance is found, the  beam energy can  be adjusted to the
peak (even if this requires a significant reduction in the luminosity)
and then the production cross section becomes very large:

\begin{equation}
\bar{\sigma}(\mu^+ \mu^- \ra
\sigma,\pi{'} ^3 \rightarrow c\bar{c})\approx {\frac{8\pi}
{M_{\Phi}^2}}\biggl({\frac{m_{\mu}^2}{3m_c^2}}\biggr) \approx 80pb  
\biggl({\frac{500GeV}{M_{\Phi}}}\biggr)^2.
\end{equation}

With  a  luminosity of  $2   \times 10^{34} cm^{-2}s^{-1}$ ($7  \times
10^{32} cm^{-2}s^{-1}$), and   a  c-tagging efficiency of  30\%,   the
observed rate should be  $10^7$  ($\sim$ $2  \times 10^5$) events  per
year.

\subsection*{BESS Model Study of SEWS}
The BESS model is an effective  Lagrangian parametrization approach to
the symmetry breaking mechanism.  The symmetry group  of the theory is
$G' =  SU(2)_L\times SU(2)_R\times SU(2)_V$, where  $SU(2)_V  $ is the
hidden symmetry through which new vector particles are introduced. The
spontaneous breakdown of  the   symmetry group  $G'\rightarrow  SU(2)$
gives  rise to six  Goldstone bosons.  Three of  these are absorbed by
new vector  particles while the other three  give mass to the SM gauge
bosons when gauging of the  subgroup $ SU(2)_L \times SU(2)_Y~\subset~
G$ is performed.

\begin{figure}[t]
\begin{minipage}[t]{2.5in}
\vspace*{7cm}
\includegraphics{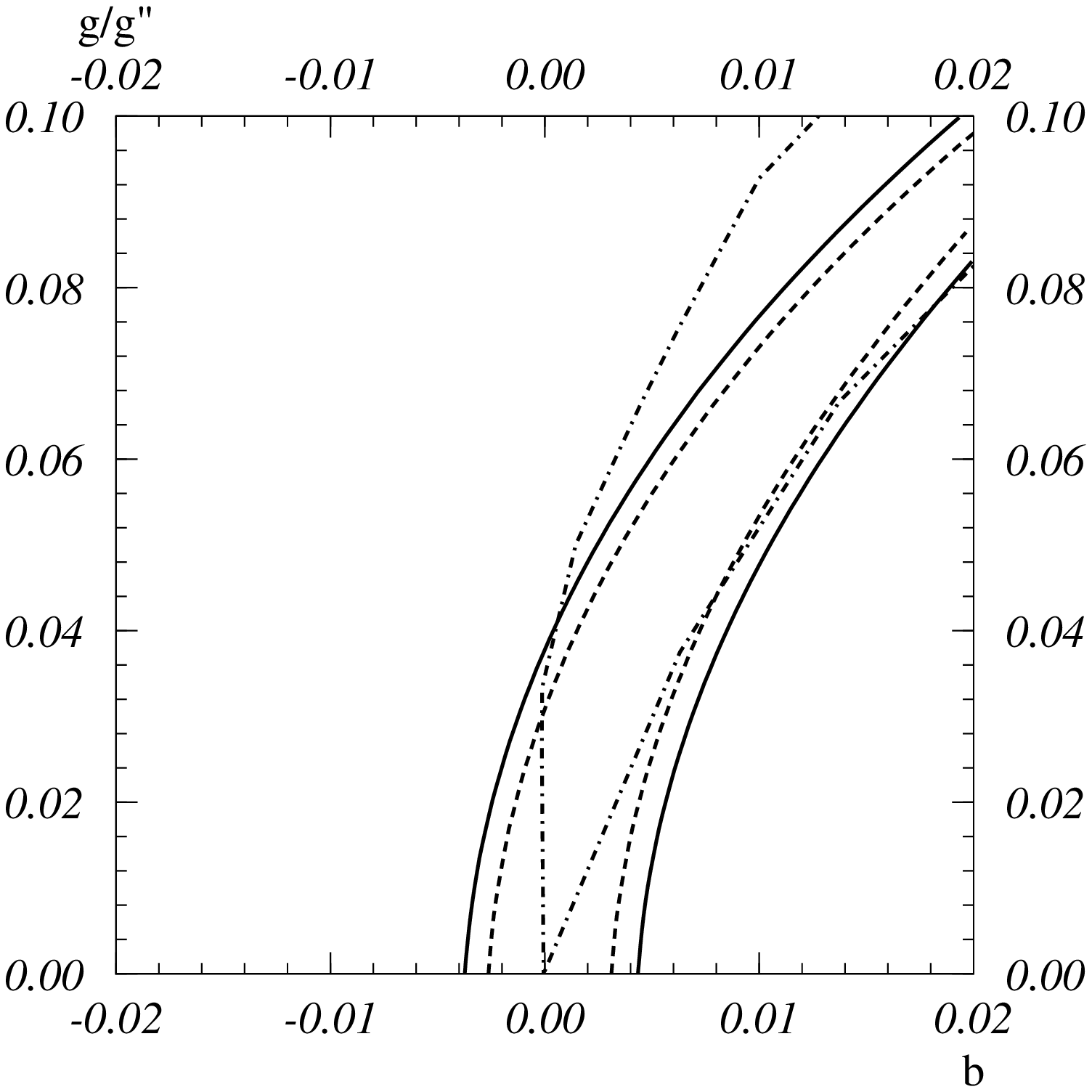}
\caption[bess]{90\% C.L. contours from BESS model, for
$M_V$=1~TeV.  The  allowed region is  between two lines. 
See text for details.}
\label{fig:dom1}
\end{minipage}\hfill
\begin{minipage}[t]{2.5in}
\vspace*{7cm}
\includegraphics{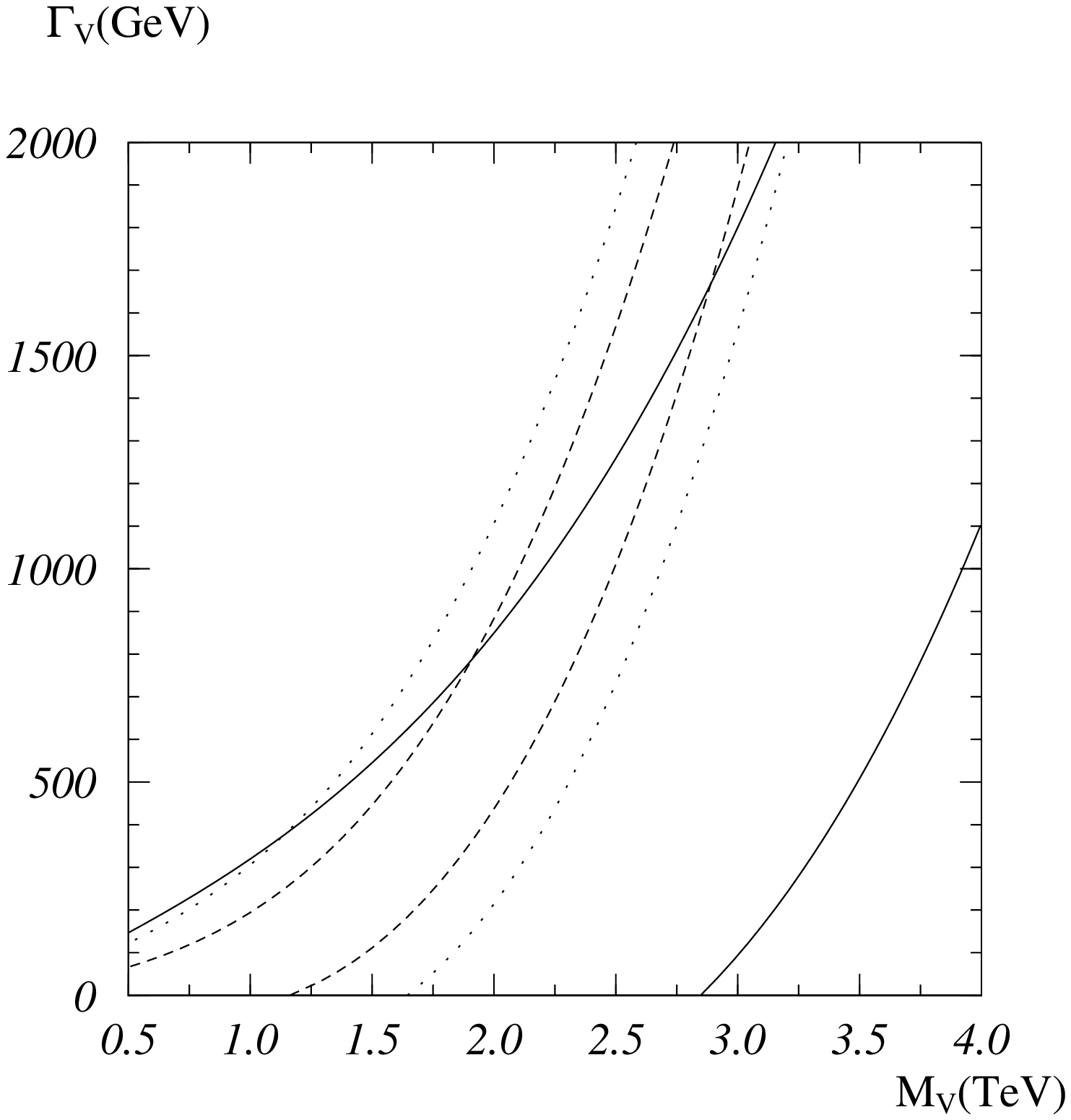}
\caption[pwu]{Partial wave unitarity bounds in the ($M_V,
\Gamma_V$) plane for $\Lambda/M_V$ = 1.5.  The dashed lines corresponds
to  the  partial wave $a_{00}$, the dotted ones to $a_{20}$  and the
solid lines to $a_{11}$.}
\label{fig:dom2}
\end{minipage}
\end{figure}

The  parameters of the BESS  model  are the masses   of the new bosons
$M_V$, their   self-coupling    $g''$,  and  a  parameter    $b$  that
characterizes the  coupling strengths of  $V$ to the  fermions. Taking
$b\rightarrow 0$ and $g''\rightarrow  \infty$, the new bosons decouple
and the SM is recovered. Bounds on  the parameter space obtained by an
analysis               of             $d\sigma(\ell^+\ell^-\rightarrow
W_{L,T}^+W_{L,T}^-)/dcos\theta$ ($\theta$  being the scattering  angle
of    the    $W$   in  the    center    of   mass),  are     shown  in
Fig.~\ref{fig:dom1}.  The solid  lines  show the  case  relevant to an
$e^+e^-$  machine with   ${\sqrt   s}$=500 GeV  and  $\int{\it L}dt=20
fb^{-1}$,  the dashed lines correspond  to a $\mu^+\mu^-$ machine with
same ${\sqrt  s}$ and luminosity.  The $\mu^+\mu^-$  collider provides
some improvement in the bounds. The result for LHC with $pp\rightarrow
W^{\pm}V^{\mp}\rightarrow W^{\pm}Z$ is  shown by  dot-dash curves  for
comparison.

Partial wave unitarity bounds from WW scattering deduced in the $(M_V,
g/g'')$ and  $(\Gamma_V, M_V)$ planes  (see Fig.~\ref{fig:dom2}) imply
that one  or more of the heavy  vector resonances should be discovered
at the   LHC, NLC or  a ${\sqrt  s}\sim$500~GeV muon  collider or, for
certain, at a  3-4~TeV muon collider, unless  $g''$ is  very large and
$b$ is very small so that they are largely decoupled.

Since the workshop, the production of  the lightest neutral PNGB  
($P^0$) in  the $s$-channel,
and  the potential  for discovering  it at  the FMC have  also been studied
using an  extension   of the model with  $SU(8)\times
SU(8)$ symmetry~\cite{dominici}. 

\subsection*{Search for Technicolor at CDF}

The  CDF collaboration reported   on their search  for technipion  and
technirho signals  in the W~+~2   jets~+~$b$-tag channel in the  Run I
data.  The signautres sought are for the processes:

$$q\bar{q} \rightarrow W^{*^\pm} \rightarrow \rho_T^{\pm} \rightarrow
 W^\pm \pi_T^0$$ \\
 and
$$q\bar{q} \rightarrow Z^*, \gamma^*
\rightarrow \rho_T^0 \rightarrow W^\pm \pi_T^\mp$$ \\
 with $W^\pm
\rightarrow \ell\nu$ ($\ell=e$ or $\mu$) and $\pi_T^0 \rightarrow b\bar{b},
~\pi_T^\pm \rightarrow b\bar{c},~ c\bar{b} ~(\approx 95\%)$ and
$\pi_T^\pm \rightarrow c\bar{s},~ s\bar{c} ~(\approx 5\%)$.

\begin{figure}[p] 
\vspace*{10cm}
\includegraphics{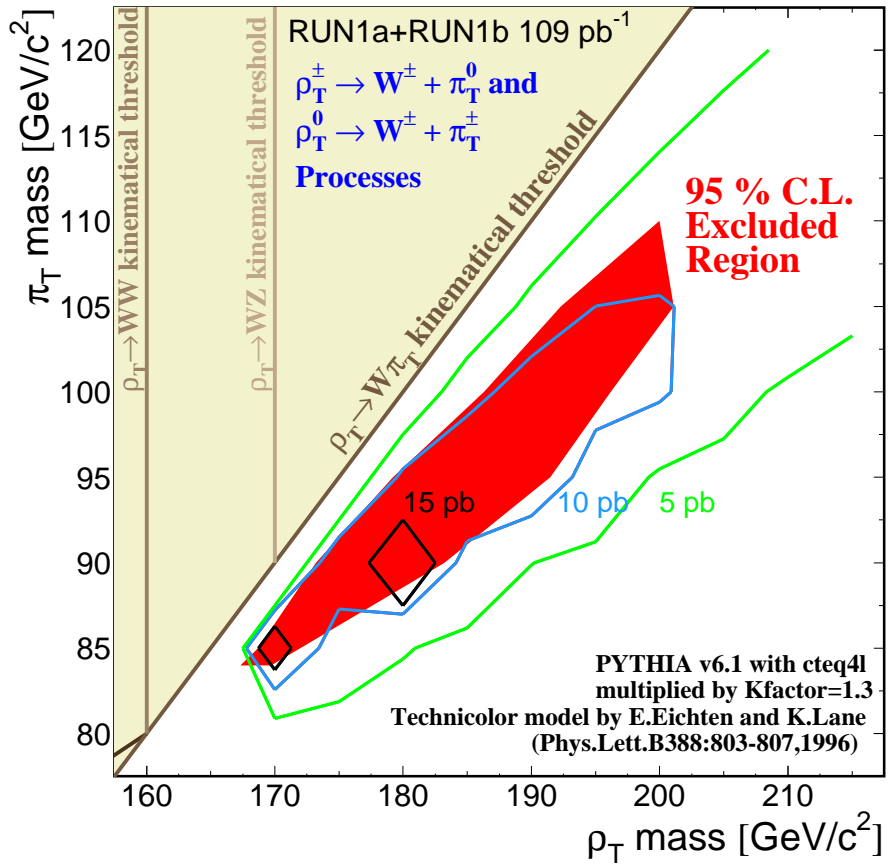}
\caption[cdf]{The 95\% C.L. exclusion region in 
(M $(\pi_T)$, M$(\rho_T)$) plane.  Some production
cross section contours are also shown.}
\label{fig:cdf}
\end{figure}

The candidate event  selection  requires an isolated electron   (muon)
with $E_T(p_T)>$  20 GeV within $|\eta| <  1.0$, \met\ $>$ 20  GeV and
two or more  jets with $E_T>$ 15  GeV.  At  least one  of  the jets is
required  to  be a $b$-jet,   tagged  by the silicon  vertex  detector
(SVX). The Z boson  candidates are rejected by requiring $|M_{ee}-M_Z|
>$15  \gevcc. A total  of 42  events are  selected while the  expected
number of background  from $Wb\bar{b},~Wc\bar{c}$, Wc, top production,
mis-tags, Z+heavy flavor amount to 31.6$\pm$4.3 events.

The  technicolor signal is   modeled using PYTHIA  MC and  GEANT-based
detector simulation. Signal  MC events are   generated at a  number of
($\pi_T,\rho_T$)  mass values.  The  combinations with  more than 5 pb
cross section are used. The technicolor model  parameters used are the
ones  from   ref~\cite{tpitev}.   Further   cuts  on   kinematic
variables $\Delta \phi(jj)$ (the azimuthal angle between two jets) and
$p_T(jj)$ ($p_T$  of   the  dijet system)~\cite{JW} are   employed  to
enhance  the  expected signal to  background    ratio in the  selected
sample. Finally, M($jj$) and M(W$jj$)  are required to be within $\pm$
3$\sigma$ of  the expected mean values  for the signal. No significant
excess is  seen  in the  data. The  95\%  C.L.  upper limits   on  the
production  cross   section then   exclude certain region   of the  (M
$(\pi_T)$, M$(\rho_T)$) plane as shown in Fig.~\ref{fig:cdf}.

\section*{Probing Muon Compositness at the  FMC}

The generational pattern  of quarks  and  leptons hints  possibly at a
substructure (with an associated  strong  interaction at energy  scale
$\Lambda  $) that might manifest at   high energies.  The existence of
such substructure, however, is    expected to result   in four-fermion
``contact'' interactions which differ  from those arising from the SM,
at energies well below  $\Lambda $.  The signals  can  be sought in  a
number   of  ways---inclusive  jet  production, Drell-Yan  production,
Bhabha scattering {\it etc}.  CELLO at the $e ^+  e ^-$ collider PETRA
with a  $\sqrt s $= 35  GeV and $\int{\cal L}dt $=   86 $pb^{-1} $ was
able to set a limit on the electron compositeness scale $\sim $2-4 TeV
using Bhabha scattering. These limits are similar to the ones from the
Tevatron ($p\bar{p},~{\sqrt s}=1.8~TeV$) ~\cite{cdfcomp}. Clearly, the
lepton colliders  seem to  hold  great  potential for probing   lepton
compositeness.  Probing the muon compositeness using Bhabha scattering
measurements  and the reach attainable as  a function of  $\sqrt s$ at
the muon   colliders  has been  investigated   by Eichten  and  Keller
~\cite{keller}.

\begin{figure}
\vspace*{6.2cm}
\includegraphics{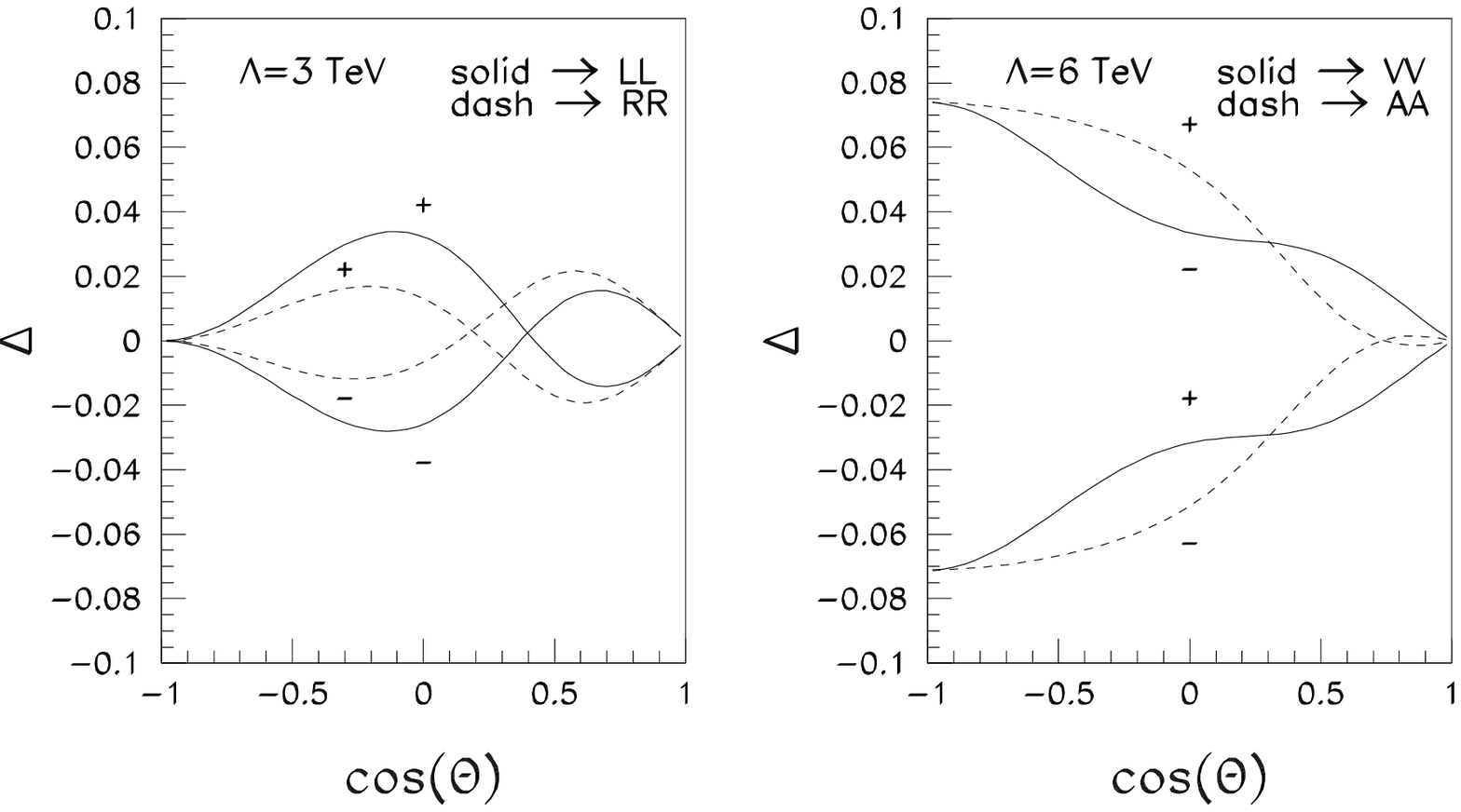}
\caption[s100]{The variable $\Delta$ versus cos$\theta$ at ${\sqrt
s}$=100~GeV for the four models, LL, RR, VV, and AA, for the two signs
of the $\eta$'s, indicate by $+$ and $-$ on the plot.}
\label{fig:plot100}
\end{figure}

\begin{figure} 
\vspace*{6.2cm}
\includegraphics{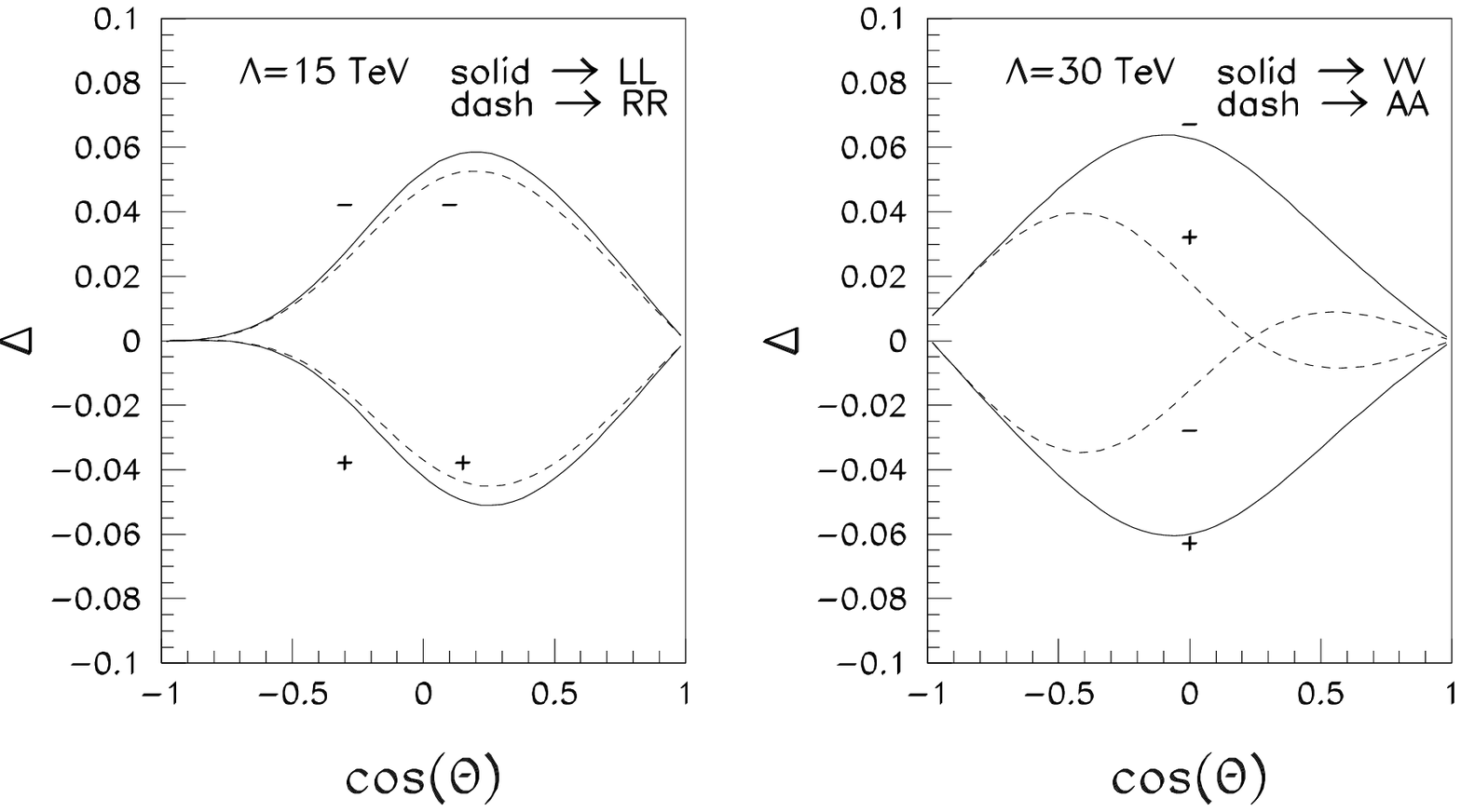}
\caption[s500]{The  variable   $\Delta$    versus  cos$\theta$ at    ${\sqrt
s}$=500~GeV for the four models, LL, RR, VV, and AA, for the two signs
of the $\eta$'s, indicate by $+$ and $-$ on the plot.}
\label{fig:plot500}
\end{figure}

The four-fermion contact interaction is assumed to be described by the
effective  Lagrangian   proposed     by Eichten, Lane      and  Peskin
\cite{composite}:
\begin{equation}
{\cal L} ={\frac {g^2}{2\Lambda^2}}[\eta_{LL}j_L j_L + \eta_{RR}j_R j_R
+\eta_{LR}j_L j_R]
\end{equation}
where $j_L$  and $j_R$ are the  left-handed and right-handed currents,
respectively; $\Lambda$ is   the   compositeness scale    and  ${\frac
{g^2}{4\pi}}$=1  is assumed (strong  coupling). The quantity $\eta$ is
used to  set  the sign  of  the  coupling i.e., $|\eta|=\pm   1$. Four
typical   coupling  scenarios are considered   in   the present work :
$LL(\eta_{LL}=\pm 1,\eta_{RR}=\eta_{LR}=0),   RR(\eta_{RR}=\pm      1,
\eta_{LL}=\eta_{LR}=0), VV(\eta_{LL}=\eta_{RR}=\eta_{LR}   =\pm1)$ and
$AA(\eta_{LL}=\eta_{RR}=-\eta_{LR}=\pm1)$. The angular distribution of
scattered muons (scattering   angle $\theta$) are  then calculated for
each of the models,  with  and without compositeness hypothesis.   The
fractional   change   in the   differential   cross   section due   to
compositeness,
\begin{equation}
\Delta=\frac {(\frac {d\sigma}{dcos\theta})_{EW+\Lambda}-
(\frac {d\sigma}{dcos\theta})_{EW}}{(\frac {d\sigma}{dcos\theta})_{EW}}
\end{equation}
is  shown in Figures.~\ref{fig:plot100}  and \ref{fig:plot500} for the
four  different models and for both  signs of $\eta$'s.  The plots are
made with $\Lambda$ chosen to  provide  an average correction of  10\%
due to compositeness.

\begin{table}
\caption[LEP]{95\% CL limits (in TeV) for various ${\sqrt s}$ (in GeV)
of  the  muon collider ($|\cos\theta|<  0.8$  required).  Expected LEP
limits ($|\cos\theta|< 0.95$) are also shown.  }
\label{tab:keller1}
\begin{tabular}{|c|c|c|c|c|c|c|c|}
      & LEP(91) & LEP(175) & 100	& 200  & 350 & 500 & 4000 \\ \hline
${\cal L} (fb^{-1})$ & .15   & .1       & .6   & 1.    & 3.   & 7.   & 450.  
\\ \hline
LL    &	4.0  & 5.8       & 4.8  & 10   & 20  & 29  & 243 \\
RR    & 3.8  & 5.7 	 & 4.9  & 10   & 19  & 28  & 228 \\
VV    & 6.9  & 12. 	 & 12   & 21   & 36  & 54  & 435 \\
AA    & 3.8  & 7.2 	 & 12   & 13   & 21  & 32  & 263 \\
\hline
\end{tabular}
\end{table}
\begin{table}
\caption[cuts]{95\% CL limits (in TeV) for different on the scattering angle  
$\theta$ cuts ($\sqrt{s}=500$~GeV, ${\cal L}=7 fb^{-1}$).}
\label{tab:keller2}
\begin{tabular}{|c|c|c|c|c|}
$|\cos  \theta| <    $   & .6&.8&.9&.95  \\ \hline    LL&26&29&31&32\\
RR&24&28&30&30\\ VV&50&54&56&57\\ AA&28&32&34&35\\
\end{tabular}
\end{table}

The 95\% C.L. limits on the compositeness scale $\Lambda$ are computed
by employing an analytical approach that approximates $\chi^2$ fitting
of ideal data to theory. The limits  extracted for various ${\sqrt s}$
of  the   muon collider and  for  various  models,   together with the
expected   limits  attainable     at    LEP  are  shown       in Table
\ref{tab:keller1}.  Since the detectors   at a muon  collider may  not
provide   coverage   down    to small      angles    due to      large
backgrounds~\cite{lebrun},  the 95\%    C.L. limits   have  also  been
extracted for different cuts on cos$~\theta$. The results are shown in
Table~\ref{tab:keller2}.  It is seen  that  the reach only improves by
10\% in going from $\cos\theta$= 0.8 to $\cos\theta$=0.95. So, it does
not seem necessary to have detector coverage to very small angles.

\section*{Constraints from Rare $B$ and $K$ Decays}
The new strong dynamics scenario for EWSB or for the origin of fermion
masses can produce sizeable effects in low energy observables at
energies much smaller than the scale of new physics. Such effects in
rare $B$ and $K$ decays have been studied by Burdman~\cite{burdman},
in the framework of an effective Lagrangian Model. These effects in
FCNC processes seem to originate from the insertion of anomalous
triple gauge boson coupling vertices and four-fermion operators.

In the four-fermion operator scenario, it has been shown that
branching ratios for $B \ra q \ell^+\ell^-, b\ra q\nu\bar{\nu}, b\ra
q\bar{q}'q'$ can have large deviations (up to a factor of $\sim$2)
from the SM expectations. However, no significant deviation is expected
in $b\ra s\gamma$ decay. The effects are very similar in rare $K$
decays such as $K^+ \ra \pi^+\nu\bar{\nu}$ and $K_L \ra
\pi^o\nu\bar{\nu}$. The effective Lagrangian approach for non-SM
couplings of fermions to gauge bosons has been examined in the topcolor
class of theories. The presence of the relatively light top-pions, and
other additional bound states, imposes severe constraints on the
topcolor models due to their potential loop effects in low energy 
observables such as $R_b$ and rare $B$ and $K$ decay rates.
These depend not only on $f_{\pi_t}$ and $m_{\pi_t}$, but
typically also on one or more elements of the quark rotation matrices
necessary to diagonalize the quark Yukawa couplings. So, it can be
shown for example, for $f_{\pi_t}\approx 120$ GeV, the effect of a
400 GeV $\pi_t$ in $b\ra s\ell^+\ell^-$ is an enhancement of more than
5\% with respect to SM expectations. Similar effects are expected to
be present in $K^+ \ra \pi^+\nu\bar{\nu}$. Thus, the measurements of
$R_b$ and the rare $B$ and $K$ decay modes can constrain strong 
dynamics models such as the topcolor model.

\section*{Summary}
We have reviewed various theories that  currently offer to explain the
breaking  of    electroweak symmetry  dynamically.    In   particular,
technicolor and related theories have been  examined in detail. Direct
searches for signals of new strong dynamics  as well as indirect tests
for existence   of    strong dynamics,   at   the  FMC,    have   been
studied. Long-range opportunities at   the high energy  muon colliders
(BMC with $\sqrt s$=3-4~TeV)  are examined. The experimental prospects
for strong dynamics at the FMC can be summarized as follows:

\begin{itemize}
\item If   low   energy  technicolor   signatures (extended    walking
technicolor, topcolor-assisted technicolor) exist, they would be found
at the Tevatron or at the LHC.  In this case, 
the first muon collider will be a
remarkable  facility  to   make   detailed  studies  and     precision
measurements.   The narrow neutral technihadrons---$\pi_T,~\rho_T$ and
$\omega_T$---would appear as spectacular resonances at the FMC ($\sqrt
s$=100--200~GeV and  energy  resolution  ${\frac  {\sigma_E}{E}}  \leq
10^{-4}$). One can  operate on the  resonance and study all the decays
and    branching  fractions  of   the    technirhos,  techniomegas and
technipions.  We emphasize that the  all-hadronic modes  would be very
difficult to study at the hadron colliders.

\item The  good  beam resolution  achievable at  the FMC  with ${\sqrt
s}\sim$ 200~GeV would enable studies of $\rho_T-\omega_T$ interference
effects in detail, using fermion-antifermion final states.

\item A variety of other models such as  technicolor with scalars and
top see-saw  model predict s-channel resonance
production  of new particles. Muon collider  has a  big advantage over
other colliders to discover and study these particles.

\item Compositeness tests  using  Bhabha  scattering give reaches   of
several tens of TeV at the FMC. At the  big muon collider, the reaches
for  95\%   confidence   level limits   on  compositeness   scale  are
unparellel,   far  exceeding   the  reaches  possible    at any  other
collider. The reaches  would be of the  order of 200-300~TeV, at which
scale we would  be probing the structure  of  the dynamics that  gives
rise to fermion masses and mixings.

\item Studies  of rare B and K decays using the  front-end of the FMC
can provide tight constraints on strong  dynamics and help distinguish
between  universal (EWSB sector) {\it vs.}    non-universal (flavor
dynamics) scenarios.

\end{itemize}

These are extremely strong physics motivations to build the first muon  
collider. A full exploration of the strong electroweak sector can be  
accomplished at the big muon collider with ${\sqrt s}$=3-4~TeV.

\end{document}